\documentclass[aps,prb,twocolumn,showpacs,amsmath,amssymb,floatfix]{revtex4}
\usepackage{epsfig}
\usepackage{bm}
\usepackage{color}

\begin{document}

\title{Scattering theory approach to bosonization of non-equilibrium mesoscopic systems }
\author{Eugene V. Sukhorukov}
\affiliation{D\'epartement de Physique Th\'eorique, Universit\'e de Gen\`eve,
CH-1211 Gen\`eve 4, Switzerland
}
\email{eugene.sukhorukov@unige.ch}

\date{\today}

\begin{abstract}
Between many prominent contributions of Markus B\"uttiker to mesoscopic physics, the scattering theory approach to the electron transport and noise stands out for its elegance, simplicity, universality, and popularity between theorists working in this field. It offers an efficient way to theoretically investigate open electron systems far from equilibrium. However, this method is limited to situations where interactions between electrons can be ignored, or considered perturbatively. Fortunately, this is the case in a broad class of metallic systems, which are commonly described by the Fermi liquid theory. Yet, there exist another broad class of electron systems of reduced dimensionality, the so-called Tomonaga-Luttinger liquids, where interactions are effectively strong and cannot be neglected even at low energies. Nevertheless, strong interactions can be accounted exactly using the bosonization technique, which utilizes the free-bosonic character of collective excitations in these systems. In the present work, we use this fact in order to develop the scattering theory approach to the bosonization of open quasi-one dimensional electron systems far from equilibrium. 

\end{abstract}


\maketitle

\section{Introduction}

Non-equilibrium phenomena in condensed matter systems are notoriously difficult to study theoretically, because one is not able to rely on the universal relations for equilibrium state, such as fluctuation-dissipation relations, and because interactions, phase coherence, quantum and many-body effects manifest themselves in entirely different way compared to the equilibrium case, requiring for their description new techniques. Many theoretical methods developed for non-equilibrium systems and phenomena can nowadays be found in textbooks, ranging from kinetic theories,\cite{LP} to the more sophisticated functional Keldysh technique.\cite{Kamenev} However, in the context of the mesoscopic physics of small metallic systems, the scattering theory approach of Markus B\"uttiker to the electron transport\cite{MB,SB} has proven to be perhaps the most efficient and widely used method. The great success of this method may be explained by the simplicity of the required calculations, and by relative universality of its applications. Indeed, this approach relies on the free-fermionic character of the electron transport, justified by the Fermi-liquid theory of metals,\cite{Abrikosov} according to which interactions can be often neglected at low energies.

However, the recent progress in the experimental techniques has revealed the new very interesting class of mesoscopic systems, quasi-one dimensional (1D) conductors, which being perhaps less broad, become nevertheless more and more experimentally accessible nowadays. Since the earlier works of Tomonaga\cite{Tomonaga} and Luttinger,\cite{Luttinger} who proposed and solved the model Hamiltonians for quasi-1D systems (dubbed as the Tomonaga-Luttinger liquids), it is known that interactions of the constituting fermions cannot be neglected even at low energies, and very often they are not perturbative, in contrast to the Fermi liquid case. Therefore, a special theoretical technique, the so-called bosonization, has been proposed to deal with interactions (for a review, see Refs.\ [\onlinecite{Gogolin}] and [\onlinecite{Thierry}]). The bosonization technique, roughly speaking,  replaces fermions with bosons of the collective excitations, such as charge density or current. This nonlinear transformation is quite complex and non-trivial.  Nevertheless, it is  relatively well understood, rigorously described, and widely used for equilibrium systems of a finite size, where it  relies on imposing periodic boundary conditions on the fields.\cite{Thierry} 

On the other hand, in the case of open quantum systems far away from equilibrium very often one faces a difficulty that periodic boundary conditions cannot be applied to the fields. This, for example, concerns the mesoscopic systems, which are attached to reservoirs of electrons (Ohmic contacts) and voltage biased in order to study an electron transport, i.e., the situation that has been considered by Markus B\"uttiker in the case of free electrons. In order to address strong interactions in non-equilibrium chiral quasi-1D systems, the Ref.\ [\onlinecite{NB}] has proposed to solve equations of motion for the bosonic fields with arbitrary boundary conditions in order to express fermion correlators in terms of the statistics of the currents of free fermions\cite{FCS} away from the scattering region. This technique has been  successfully used to explain recent experiments with quantum Hall (QH) edge states.\cite{experiments} However, the free-bosonic character of excitations in quasi-1D systems of fermions with a linearized spectrum suggests that perhaps as simple and powerful scattering theory as the one of Markus B\"uttiker can also be formulated in the case of strong interactions. Our work presents an effort in this direction.

Earlier versions of the scattering theory for bosons in quasi-1D electron systems have been proposed in various context, including inter-edge interactions in QH systems,\cite{Oreg} the universality of the DC conductance of quantum wires,\cite{Ines1} thermal transport,\cite{Fazio} the frequency-dependent linear response,\cite{Ines2} resonant dephasing in electronic interferometers,\cite{SC} energy exchange at the QH edge,\cite{Degiovanni}  equilibration of QH edge states by an Ohmic contact,\cite{us2} and the decoherence of single-particle excitations
at the QH edge.\cite{Dario} Here, instead of focusing on particular physical phenomena,  we formulate the scattering theory approach to the bosonization on a more rigorous  level. The goal of this approach is to overcome limitations of the scattering theory for free fermions by accounting for a broad class of strong density-density interactions non-perturbatively. The trade-off of this technique is that the fermion ``mixing'', i.e., electron tunneling and backscattering effects have to be taken into account perturbatively\cite{footnote0} (with an exception of the boundary conditions considered in Sec.\ \ref{boundary-c}, and of the electron mixing in reservoirs discussed in Sec.\ \ref{LE}). This requires the knowledge of electron correlation functions. Our present work proposes a framework for the calculation of such correlators for open quasi-1D electronic systems.  

The rest of the paper is organized as follows. We start in Sec.\ \ref{IB} with the pedagogical introduction to the bosonization, introduce interactions and formulate the scattering problem for bosons. This is followed in Sec.\ \ref{SS} by the proof of the orthogonality and completeness of the basis of scattering states. This step is used to prove the fermionic commutation relations for vertex operators, and thus completes the bosonization of the interacting fermions. Zero modes in open systems acquire a new physical meaning and properties, which are the subject of Sec.\ \ref{ZM}. We proceed in Sec.\ \ref{boundary-c} with arbitrary boundary conditions for the fields, and connect fermionic correlators to the full counting statistics (FCS) of free-fermionic currents,\cite{FCS} thereby generalizing the results of Ref.\ [\onlinecite{NB}]. Finally, in Sec.\ \ref{LE} we combine the scattering theory with the quantum Langevin equations in order to account for the effects of dissipation and fluctuations, arising in the electrical circuit, to which a quasi-1D system is attached.

\section{Introduction to bosonization}
\label{IB}

We start with the simple example of free chiral fermions, representing electrons in a quasi-1D mesoscopic system, and add interactions below in
this section.
The details of the bosonization procedure may be found in a number of textbooks,
for example, in Ref.\ [\onlinecite{Thierry}]. The formulation of the problem in the context 
of chiral quantum Hall edge states, both at integer and fractional filling factors,
may be found, for instance, in Refs.\ [\onlinecite{ourwork1}] and [\onlinecite{ourwork2}]. In this section we outline, in a pedagogical manner,  only essential steps needed for understanding the rest of the paper. Throughout the paper we use unites, where $e=\hbar=k_B=1$. 

\begin{figure}[h]
\epsfxsize=7cm
\epsfbox{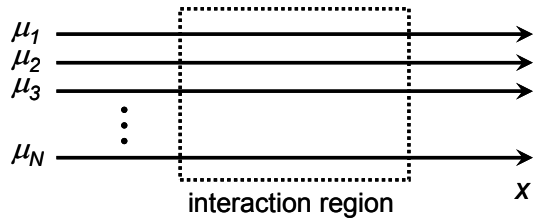}
\caption{Schematic representation of a quasi-1D mesoscopic system, consisting of a set of chiral fermions,
interacting in a finite region confined inside the dotted box. Although,
the spatial location and orientation of the channels may be arbitrary,
we choose, without loss of the generality of our analysis, the same parametrization with the coordinate $x$. 
Each channel may originate from its own reservoir, biased with 
the electro-chemical potential $\mu_n$, and having its own temperature $T_n$, 
$n=1,\ldots,N$.  
} \vspace{-3mm}
\label{schematics1}
\end{figure}

\subsection{Free fermions}

Let us consider a system of $N$ chiral fermions $\psi_n(x)$, 
where $n=1,\ldots,N$, originating from their own reservoirs with the temperatures $T_n$,  biased with the 
electro-chemical potentials $\mu_n$. The spatial location and orientation of 
each channel may be arbitrary. It is only for the convenience, and without loss of the generality, we parametrize the channels
with the same coordinate $x$, as shown in Fig.\ \ref{schematics1}. The fermions carry the charge density:
\begin{equation}
\rho_n(x)=\psi_n^\dagger(x)\psi_n(x).
\label{charge-def}
\end{equation}
Using the standard commutation relations for fermions
\begin{equation}
\{\psi_n(x),\psi_{n'}^\dagger(x')\}=\delta_{nn'}\delta(x-x'),
\label{fermion-com}
\end{equation}
one obtains the relation of the charge locality
\begin{equation}
[\rho_n(x),\psi_{n'}^\dagger(x')]=\delta_{nn'}\delta(x-x')\psi_{n}^\dagger(x),
\label{locality}
\end{equation}
which imply that the operator $\psi_{n'}^\dagger(x')$, representing an electron, creates the charge equal to 1 in the $n$th channel at the point $x=x'$.

The Hamiltonian of free chiral fermions with the spectrum linearized in the vicinity of the Fermi level
reads
\begin{equation}
H_0=-i\sum_n v_n\!\int dx\psi^\dagger_n(x)\partial_x\psi_n(x),
\label{free-h}
\end{equation}
where $v_n$ are the group velocities of fermions at the Fermi level. The equations of motion $\partial_t\psi_n=i[H_0,\psi_n]$, which immediately follow from the commutation relations (\ref{fermion-com}),
\begin{equation}
(\partial_t+v_n\partial_x)\psi_n(x,t)=0,
\label{eq-motion}
\end{equation}
describe chiral waves that propagate with constant speeds.

In equilibrium, one imposes periodic boundary conditions, $\psi_n(x,t)=\psi_n(x+L_n,t)$, where $L_n$ is the size of the {\em n}th channel, and presents the solution as a sum over plane waves. Taking the thermodynamic limit, $L_n\to\infty$, the solution reads
\begin{equation}
\psi_n(x,t)=\frac{1}{\sqrt{2\pi v_n}}\int\limits_{-\infty}^\infty\! d\omega\, e^{-i\omega\tau}c_n(\omega) ,\;\; \tau\equiv t-x/v_n,
\label{decomposition}
\end{equation}
where
\begin{equation}
\{c_n(\omega),c_{n'}^\dagger(\omega')\}=\delta_{nn'}\delta(\omega-\omega').
\label{canonical}
\end{equation}
It is easy to see, that the operators (\ref{decomposition}) satisfy the commutation relations (\ref{fermion-com}).

Finally, we note that at ground state 
$\langle c_n(\omega)c_{n}^\dagger(\omega')\rangle=\delta(\omega-\omega')\theta(\omega-\mu_n)$,
i.e., all the states below the Fermi level are occupied, while all the states above the Fermi level are empty. Therefore, the free-fermionic correlation function
reads
\begin{equation}
\langle \psi_n(x,t)\psi_{n}^\dagger(x',t')\rangle=\frac{i}{2\pi v_n}\frac{e^{i\mu_n(\tau'-\tau)}}{\tau'-\tau+i0}.
\label{corr-free}
\end{equation}
We will use this result below as a reference.

\subsection{Free bosons}

\begin{figure}[h]
\epsfxsize=7cm
\epsfbox{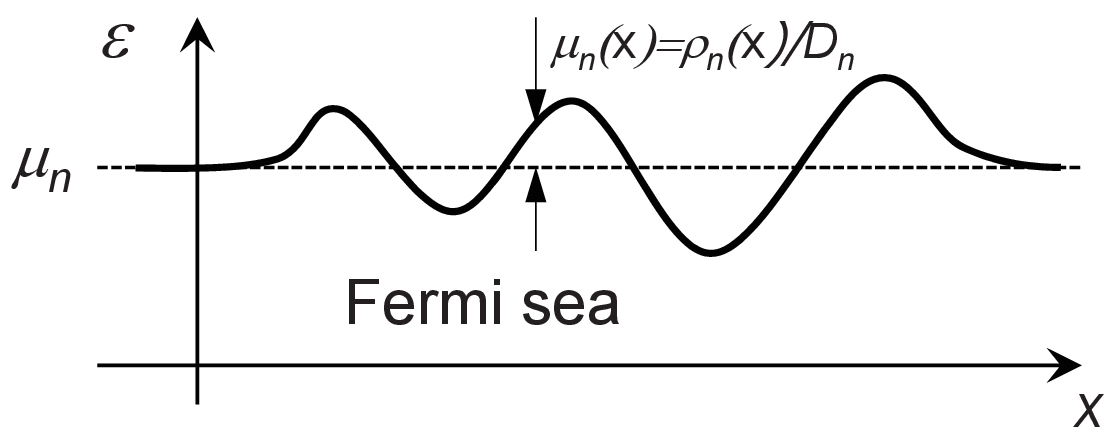}
\caption{ Deformation of the Fermi sea in one of the channels is schematically shown. All the states below the wavy line are filled, while all the states above this line are empty. This deformation leads to the accumulation of the coordinate-dependent 1D charge density $\rho_n(x)$.
} \vspace{-3mm}
\label{schematics2}
\end{figure}

Under the same assumption, that the spectrum of free fermions can be linearized in the vicinity of the Fermi level,
namely, if the perturbations that drive the system away from the equilibrium are relatively weak, $\mu_n,T_n\ll \varepsilon_F$, where $\varepsilon_F$ is the Fermi energy, one may consider such perturbations as incompressible deformations of the Fermi sea (shown schematically in Fig.\ \ref{schematics2}). 
The simplest example is given by free fermions with the density accumulated as a result of the shift of the electro-chemical potential: 
\begin{equation}
\langle\rho_n\rangle=D_n\mu_n,\quad D_n=1/(2\pi v_n),
\label{density-of-states}
\end{equation}
 where $D_n$ is the density of states of the $n$th channel at Fermi level.
This relation leads to the well-known universal expression for the 1D charge current, $\langle j_n\rangle=v_n\langle \rho_n\rangle= \mu_n/2\pi$. 

In the next step, we assume that these relations also hold locally for coordinate-dependent deformations, and expand the grand potential of a quasi-1D system of free fermions to second order in small deformations of the Fermi sea,
\begin{equation}
E=\sum_n\frac{1}{2D_n}\int dx \rho_n^2(x)-\mu_nQ_n, 
\label{deformations}
\end{equation}
where $Q_n=\int dx \rho_n(x)$ is the total charge in the {\em n}th channel.
Varying densities, $\delta E/\delta\rho_n=0$, while keeping potentials $\mu_n$ constant, one obtains the relation (\ref{density-of-states}) for the ground state. On the other hand, according to Eq.\ (\ref{eq-motion}), all fermions move with the constant speeds.
Therefore, the densities should satisfy same equations:
\begin{equation}
(\partial_t+v_n\partial_x)\rho_n(x,t)=0.
\label{eq-motion2}
\end{equation}
However, if the densities are replaced by operators, and $E$ is regarded as a Hamiltonian,
 the same equations should follow from the equation of motion $\partial_t \rho_n=i[E,\rho_n]$. This is possible
only if the densities satisfy the following commutation relations:
\begin{equation}
[\rho_n(x),\rho_{n'}(x')]=-(i/2\pi)\delta_{nn'}\partial_x\delta(x-x').
\label{comm-dens}
\end{equation}
This completes the quantization of the deformations of the Fermi sea.

It is convenient, and quite common, to replace the densities and currents with the displacement  fields
$\phi_n$:
\begin{equation}
\rho_n(x)=\frac{1}{2\pi}\partial_x\phi_n(x),\quad j_n(x)=-\frac{1}{2\pi}\partial_t\phi_n(x),
\label{densities}
\end{equation} 
where the last equation follows from the charge continuity relation.
Then, the commutation relations read
\begin{equation}
[\phi_n(x),\phi_{n'}(x')]=i\pi\delta_{nn'}{\rm sgn}(x-x').
\label{commutators}
\end{equation} 
We proceed, as for free fermions, by solving equations of motion in terms of plane waves, imposing periodic boundary conditions, and taking the thermodynamic limit. The resulting spectral decomposition for the new fields reads
\begin{equation}
\phi_n(x,t)=\varphi_n(\tau)
+\int\limits_0^\infty\frac{d\omega}{\sqrt{\omega}}
\left[e^{-i\omega\tau}a_n(\omega)+\mbox{h.c.}\right],
\label{fields1}
\end{equation}
where the creation and annihilation operators satisfy bosonic
commutation relations
\begin{equation}
[a_n(\omega),a^\dagger_{n'}(\omega')]=\delta_{nn'}\delta(\omega-\omega'),
\label{commutators-a}
\end{equation}
and the zero modes, 
\begin{equation}
\varphi_n(\tau)=-\varphi^{(0)}_n-2\pi v_nQ_n\tau/L_n,
\label{zm1}
\end{equation}
account for the homogeneous part of the charge density,
and change the number of fermions $Q_n$ in each channel by $1$, 
i.e., $[Q_n,e^{i\varphi^{(0)}_n}]=1$. 

Formally, the zero modes ensure correct commutation relations for the fields $\phi_n$, as well
as periodic boundary conditions. However, in the thermodynamic limit, $L_n\to\infty$, and 
for an open system, their quantum nature may typically be ignored (see, however, the discussion in Secs.\  \ref{ZM} and \ref{LE}). Indeed, at any finite distances $\Delta x$, the quantum fluctuation of the zero mode scales as
$2\pi\Delta Q_n\Delta x/L_n\sim \Delta x/L_n\to 0$.
Therefore, using Eq.\ (\ref{density-of-states}) one can replace zero modes 
with their average values:
\begin{equation}
\varphi_n(\tau)\to\langle \varphi_n(\tau)\rangle=-\mu_n\tau,
\label{0-modes1}
\end{equation}
where, we recall, $\tau\equiv t-x/v_n$.
For instance, according to Eq.\ (\ref{densities}), the average stationary currents 
acquire the values $\langle j_n\rangle=-\partial_t \varphi_n(\tau)/2\pi=\mu_n/2\pi$, 
in full agreement with the Fermi liquid theory.

\subsection{Bosonization of fermions}
\label{BofF}

The two alternative descriptions of the chiral fermions outlined above are unified by  the bosonization
procedure,\cite{Thierry} according to which the fermions are expressed in terms of bosons 
as follows
\begin{equation}
\psi_n(x)\propto e^{i\phi_n(x)}, \quad n=1,\ldots,N,
\label{def}
\end{equation}  
where the prefactor constants are determined by the high-energy cut-off
and can be found in the end of calculations by, e.g., comparing the correlators of so defined fermions to the correlation functions (\ref{corr-free}).
Using the spectral decomposition (\ref{fields1}), one can verify that the operators (\ref{def}) satisfy the fermionic
commutation relations (\ref{fermion-com}) and the charge locality relation (\ref{locality}).\cite{footnote1}
The subtle step of connecting the densities (\ref{charge-def}) and (\ref{densities}), as well as the Hamiltonians
(\ref{deformations}) and (\ref{free-h}), relies on the point-splitting procedure,\cite{Thierry} which accurately subtracts the contribution of the Fermi sea. The recent work [\onlinecite{us}] points to some physical aspects of this procedure in the context of the evaluation of the electron correlation functions and arrives at the conclusion that for the contributions close to the Fermi level this procedure can be ignored.

Finally, we note that the correlation functions   (\ref{corr-free}) of free fermions in the ground state may now be  obtained by using the operators (\ref{def}), substituting the spectral decomposition of the bosonic fields (\ref{fields1}), and applying the normal ordering procedure.  Importantly, the commutation relation for zero modes, $[Q_n,e^{i\varphi^{(0)}_n}]=1$, gaurantees the independence of fermions that belong to different 1D systems of finite size $L_n$:  $\langle \psi_n\psi_{n'}^\dagger\rangle=0$ for $n\neq n'$. This follows from the orthogonality of quantum states of finite 1D systems containing different number of fermions. However, for open quasi-1D fermionic systems connected to reservoirs this argument has to be reformulated, which is discussed in the Sec.\  \ref{LE}.

\subsection{Interactions}
\label{interactions}

Rewriting the total energy of free fermions in Eq.\ (\ref{deformations}) in terms of bosonic fields,
one obtains the bare Hamiltonian
\begin{equation}
H_0=\frac{1}{4\pi}\sum_n v_n\!\int dx\,[\partial_x\phi_n(x)]^2.
\label{free-h}
\end{equation}
We are interested in the situation, where the interactions between fermions
are present in the finite region of space, as schematically shown in Fig.\ \ref{schematics1}. 
We wish to consider strong Coulomb interactions. However, in real systems Coulomb interactions are
screened in a quite complex way.  Therefore, in order to keep generality of the following analysis, we present the interaction part of the Hamiltonian in the general form
\begin{equation}
H_1=\frac{1}{8\pi^2}\sum_{n,n'}\int\!\!\int dxdy U_{nn'}(x,y)\partial_x\phi_n(x)\partial_y\phi_{n'}(y),
\label{free-h}
\end{equation}
where the kernel $U_{nn'}$ is an arbitrary potential of the density-density interaction, which accounts spatial separation of channels and screening effects.  

The equations of motion for the bosonic fields, $\partial_t\phi_n=i[H_0+H_1,\phi_n]$,
immediately follow from the commutation relations (\ref{commutators}),
\begin{equation}
(\partial_t+v_n\partial_x)\phi_n(x,t)=
-\frac{1}{2\pi}\sum_{n'}\int dy U_{nn'}(x,y)\partial_y\phi_{n'}(y,t).
\label{eq-mot}
\end{equation}
These equations may be accompanied with the non-trivial boundary conditions, as described 
in Sec.\ \ref{boundary-c}, and solved directly. However, as a first step, we  wish to express fields
in the second-quantized form. 

We have done so for free-fermion case, 
$U_{nn'}=0$, where the fields can be expanded in terms of plane waves, see Eq.\  (\ref{fields1}).
Taking into account our assumption, that the interaction is localized in the finite region of space,
we look for the solution of the equations (\ref{eq-mot}) in the form
\begin{equation}
\phi_n(x,t)=\varphi_n(x,t)+\delta\phi_n(x,t).
\label{splitting}
\end{equation}
where the zero mode contribution reads
\begin{equation}
\varphi_n(x,t)=-\mu_n t+\tilde \varphi_n(x)
\label{zm-representation}
\end{equation}
and the fluctuating part is decomposed in the oscillator modes as
\begin{equation}
\delta\phi_n(x,t)=\int\limits_0^\infty\frac{d\omega}{\sqrt{\omega}}
\sum_m\left[\Phi_{mn\omega}(x)e^{-i\omega t}a_m(\omega)+\mbox{h.c.}\right].
\label{fields2}
\end{equation}
Here, the operators $a_m(\omega)$ are defined in Eq.\ (\ref{commutators-a}),
and the wave functions $\Phi_{mn\omega}(x)$ are the scattering states, which
acquire the form of the plane waves far away from the scattering region. In the next section,
we develop the formal scattering theory and prove the orthogonality and completeness of the
scattering states in order to guarantee the bosonic commutation relations (\ref{commutators}) 
of the operators (\ref{fields2}). The zero modes $\varphi_n(x,t)$ deserve a special consideration, which is done 
in Sec.\ \ref{ZM}. 

Finally, assuming the Gaussian character of fluctuations, one can expand fermionic operators (\ref{def}) to second order in the fields $\delta\phi_n$, average over the bosonic states, and re-exponentiate the result. This procedure leads to the following expression:
\begin{multline}
\ln \langle \psi_n(x,t)\psi_{n}^\dagger(x',t')\rangle=-i\mu_n(t-t')\\
+i[\tilde \varphi_n(x)-\tilde \varphi_n(x')]+G_n(x,x',t-t').
\label{FR1}
\end{multline}
Here, first term on the righ hand side may be attributed to the energy shift due to the applied chemical potential $\mu_n$, while the second term is the Friedel phase shift induced by the interaction, since 
$\partial_x\tilde \varphi_n(x)=2\pi\langle\rho(x)\rangle$. The fluctuation contribution
\begin{equation}
G_n(x,x',t)=\langle[\delta\phi_n(x,t)-\delta\phi_n(x,0)]\delta\phi_n(x',0)\rangle 
\label{FR2}
\end{equation}
may be found by substituting the spectral decomposition (\ref{fields2}) and taking the principle value of the integral over $\omega$, because the $\omega=0$ contribution has been attributed to the zero mode.

\section{Scattering states}
\label{SS}

In this section we formulate the scattering theory based 
on equations (\ref{eq-mot}). Namely, we introduce scattering states 
and prove their orthogonality and completeness. In our case, the scattering 
states are represented by $N$ sets of functions, $\Phi_{mn\omega}(x)$,
$m,n=1,\ldots,N$, where the first index enumerates the sets, and the second index enumerates
the functions in a particular set. They satisfy the equations of motion (\ref{eq-mot}), which  in the Fourier space
read
\begin{multline}
(i\omega-v_n\partial_x)\Phi_{mn\omega}(x)\\ =\frac{1}{2\pi}\sum_{n'}\int dy U_{nn'}(x,y)\partial_y\Phi_{mn'\omega}(y),
\label{scat-st1}  
\end{multline}
and the following boundary conditions at $|x|\to\infty$,
\begin{equation}
\Phi_{mn\omega}(x) =[\delta_{mn}\theta(-x)+S_{nm}(\omega)\theta(x)]\,e^{ik_nx},
\label{boundary-cond}  
\end{equation}
where $k_n=\omega/v_n$, and $S_{nm}$ are the elements of the scattering matrix. In other words,
the $m$th scattering state describes one incoming mode in the $m$th channel, and outgoing modes in all $N$ channels.

As a first step, we wish to derive a useful formal relation for scattering states. 
Let us consider the sets of functions $\Phi^{*}_{mn\omega}(x)$, 
which satisfy Eqs.\ (\ref{scat-st1}) with $\omega$ on the left hand side replaced 
by $-\omega$. We multiply equations of motion for the functions $\Phi^{*}_{mn\omega}(x)$
by $\partial_x\Phi_{ln\omega'}(x)$, and equations for the functions $\Phi_{ln\omega'}(x)$ 
by $\partial_x\Phi^{*}_{mn\omega}(x)$, integrate over $x$ and sum over $n$, and subtract 
one result from another. Due to the symmetry of the interaction potential, 
$U_{nn'}(x,y)=U_{n'n}(y,x)$, the right hand side of the equations cancels, and we arrive
at the following equation: 
$\int dx\sum_n(\omega\partial_x\Phi_{ln\omega'}\Phi^{*}_{mn\omega}+\omega'
\Phi_{ln\omega'}\partial_x\Phi^{*}_{mn\omega})=0$. It is convenient to cut this 
integral at large distances, $|x|=W$, beyond the interaction region, and integrate 
by parts. Using the asymptotic form (\ref{boundary-cond}), we obtain:
\begin{multline}
(\omega'-\omega)\!\int\limits_{-W}^W dx \sum_n\Phi_{ln\omega'}\partial_x\Phi^{*}_{mn\omega}
=\omega\delta_{lm}e^{-i(k'_m-k_m)W}\\
-\omega\sum_nS_{nl}(\omega')S^*_{nm}(\omega)e^{i(k'_n-k_n)W}.
\label{relation}
\end{multline}
We note that, by choosing in this equation $\omega'=\omega$, we immediately arrive at the 
unitarity of the scattering matrix, $\sum_nS_{nl}S^*_{nm}=\delta_{lm}$.

Next, we extend the integral in Eq.\ (\ref{relation}) to infinity: $W\to\infty$. Then, for $\omega'\neq\omega$
the right hand side of the equation is fast oscillating function, which vanishes upon coarse graining,
leading to the orthogonality of scattering states. On the other hand, a care has to be taken when $\omega'$
approaches $\omega$. In this case, we may rely on the unitarity of scattering matrix to arrive at the following expression
\begin{multline}
\int\limits_{-W}^W dx \sum_n\Phi_{ln\omega'}\partial_x\Phi^{*}_{mn\omega}
=-\omega\delta_{lm}\\ \times\frac{e^{i(k'_m-k_m)W}-e^{-i(k'_m-k_m)W}}{\omega'-\omega}\,.
\end{multline}
As $W\to\infty$, the last term in this equation becomes a $\delta$-function, and we arrive at the 
orthogonality relation:
\begin{equation}
\int\limits_{-\infty}^\infty dx \sum_n\Phi_{ln\omega'}\partial_x\Phi^{*}_{mn\omega}
=-2\pi i \omega\delta_{lm}\delta(\omega'-\omega)\,.
\label{orthogonality}
\end{equation}

In order to prove the completeness of scattering states, let us multiply Eq.\ (\ref{orthogonality})
by the function $\Phi_{mn'\omega}(x')$, sum over $m$, and integrate over $\omega$. The result reads
\begin{multline}
\int\limits_{-\infty}^\infty dx\! \sum_n \Phi_{ln\omega'}(x)\bigg[
\sum_m\int\limits_{0}^\infty\frac{d\omega}{\omega}\,\partial_x \Phi^*_{mn\omega}(x)\Phi_{mn'\omega}(x')\bigg]\\
=-2\pi i \Phi_{ln'\omega'}(x'),
\end{multline}
i.e., the expression in square brackets is the unity operator in the space of chiral 
scattering states.
Similarly, multiplying Eq.\ (\ref{orthogonality}) by the function $\Phi^*_{mn'\omega}(x')$
and repeating the above steps, we arrive at the analogous expression for the unity operator in the 
space of the conjugated (anti-chiral) states. By combining these two expressions, we obtain the completeness 
relation:
\begin{multline}\frac{1}{2\pi i}
\sum_m\int\limits_{0}^\infty\frac{d\omega}{\omega}\big[\Phi^*_{mn'\omega}(x')\partial_x \Phi_{mn\omega}(x)\\
-\Phi_{mn'\omega}(x')\partial_x \Phi^*_{mn\omega}(x)
\big]
=\delta_{nn'}\delta(x-x').
\label{completeness}
\end{multline}
One may easily check this relation with the simple example  of $U_{nn'}=0$, where 
$\Phi_{mn\omega}(x)=\delta_{mn}e^{ik_nx}$.

To conclude this section, one of our main results is that, as one can easily check, the commutation relations (\ref{commutators}) for the fields (\ref{fields2}) follow from the relations of the completeness (\ref{completeness}). One can extend the results of this section to the case, where $M$ of $N$ available channels are compactified on the circles of finite size $L_n$ with the periodic boundary conditions for the fields. This limits the number of scattering states to $N-M$ remaining unconfined channels. The compactified channels do not contribute to the right hand side of the equation (\ref{relation}), because the boundary terms cancel after integration by parts. On the other hand, on the left hand side of the orthogonality relation (\ref{orthogonality}) the sum over $n$ and integral over $x$ split in two parts corresponding to the two sorts of channels. Finally, the completeness relation (\ref{completeness}) does not change its form, however, the sum runs over scattering states, i.e., from $n=M+1$ to $n=N$.

\section{Zero modes}
\label{ZM}

Let us recall, that we consider an open system, therefore zero modes in most cases may be 
considered classical fields, which satisfy equations of motion (\ref{eq-mot}). 
Below, however, we will quantize zero modes in order to account for the quantum effects
of a circuit, to which the system is attached.  Away from the scattering 
region, $x\to-\infty$, zero modes obviously acquire the form (\ref{0-modes1}) 
found earlier for a translationary invariant system. Therefore, we are looking for the 
solution in the form (\ref{zm-representation}), where
the coordinate-dependent term satisfies the equation
\begin{equation}
v_n\partial_x\tilde\varphi_n(x)+
\frac{1}{2\pi }\sum_{n'}\int dy U_{nn'}(x,y)\partial_y\tilde\varphi_{n'}(y)
=\mu_n.
\label{0-eq-mot}
\end{equation}
This equation may be interpreted as a condition of constant potential at 
the channel $n$, where the first term on the left hand side is the contribution
from the finite compressibility of the Fermi sea (Thomas-Fermi correction), while the second term results from
the Coulomb interactions. 

Such electrostatic problem can be formally solved with the help of the so called ``characteristic'' potentials\cite{Buttiker2,Sukhorukov}  $f_{mn}(x)$. Namely, one can write zero modes 
in the form:
\begin{equation}
\varphi_n(x,t)=-\mu_nt+\sum_m (\mu_m/v_m) f_{mn}(x),  
\label{0-modes2}
\end{equation}
where the characteristic potentials satisfy the following equations
\begin{equation}
\partial_x f_{mn}(x)+
\frac{1}{2\pi v_n}\sum_{n'}\int dy U_{nn'}(x,y)\partial_y f_{mn'}(y)=\delta_{mn}.
\label{char}
\end{equation}
The boundary conditions may be fixed with the help of Eq.\ (\ref{0-modes1}), 
so that the asymptotic forms read:
\begin{equation}
f_{mn}(x)=x\delta_{mn}+\theta(x)\Delta f_{mn}, \quad \mbox{at $|x|\to\infty$},
\label{0-boundary}
\end{equation}
where $\Delta f_{mn}$ are the interaction-induced phase shifts.

The equations (\ref{char}) for the characteristic potentials may be solved 
directly, e.g., perturbatively with respect to the potentials  $U_{nn'}$. 
However, if scattering states $\Phi_{mn\omega}(x)$ are already known, one can, 
alternatively, extract characteristic potentials by evaluating the limit:
\begin{equation}
f_{mn}(x)=\lim_{k_m\to 0}\frac{\Phi_{mn\omega}(x)-\delta_{mn}}{ik_m}.
\label{limit1}
\end{equation}
This simply follows from the fact that the expression on the right hand side 
of this equation satisfies the equations (\ref{char}) for the characteristic
potentials, which can be easily seen from Eqs.\ (\ref{scat-st1}). Moreover,
in the low-frequency limit and at $x\to-\infty$ we find 
$[\Phi_{mn\omega}(x)-\delta_{mn}]/ik_m=\delta_{mn}(e^{ik_mx}-1)/ik_m\to x\delta_{mn}$,
i.e., the expression (\ref{limit1}) satisfies the boundary conditions for the 
characteristic potentials. Finally, according to the
Eq.\ (\ref{boundary-cond}), the phase shifts may be found from the scattering 
matrix:
\begin{equation}
\Delta f_{mn}=\lim_{k_m\to 0}\frac{S_{mn}(\omega)-\delta_{mn}}{ik_m}.
\label{limit2}
\end{equation}
We stress that, in contrast to one-dimensional low-energy 
fermions, the bosons do not scatter at low frequencies, i.e., $S_{mn}(0)=\delta_{mn}$,
and the limit (\ref{limit2}) is well defined. 

The equations (\ref{0-modes2}) for zero modes and (\ref{char}) for the characteristic
potentials may be used in order to evaluate voltage dependent and interaction
induced specific phase shifts, which contribute to electron correlation functions.
Such phase shifts are relevant for a number of experimental situations, as has been 
demonstrated in Ref.\ [\onlinecite{ourwork1}]. However, so far we have considered
an ideal situation, where the system is attached to an electrical circuit via
voltage biased ohmic reservoirs. Sometimes, one is not able to neglect effects 
of fluctuations inside an electrical circuit, which may propagate down to the system 
along the 1D channels. A formal way to account for these effects is to impose apropriate 
boundary conditions on the boson fields, as we demonstrate in the section \ref{LE}. 
However, in the case, where the characteristic frequency of circuit fluctuations, $\omega_0$,
is much smaller than the the inverse time 
of the propagation of fluctuations through the system, $1/t_f$, there is an alternative way to proceed. 

To address this specific situation, let us consider a low-frequency limit of the oscillator 
part of the fields $\phi_n$ in Eq.\ (\ref{fields2}), which accounts for the dynamical effects.
By using the equation (\ref{limit1}), we arrive at the following expression
\begin{equation}
\varphi_n(x,t)=-\theta_n(t)+\sum_m v_m^{-1}f_{mn}(x)\partial_t\theta_m(t),
\label{circuit1}
\end{equation}
where we have introduces the operator
\begin{equation}
\theta_n(t)=-\int\limits_{0}^{\omega_c}\frac{d\omega}{\sqrt{\omega}}
\left[e^{-i\omega t}a_n(\omega)+\mbox{h.c.}\right],
\end{equation} 
and $\omega_c$ is the energy cutoff, such as $\omega_0\ll\omega_c\ll 1/t_f$.
Note, that $\langle\partial_t\theta_n\rangle=\mu_n$, and if the fluctuations 
can be neglected, we come back to the equation (\ref{0-modes2}). In order to take into
account fluctuations, the operator $\partial_t \theta_n$ has to be considered a circuit variable,
namely, a fluctuating potential in reservoirs. Note also, that Eq.\ (\ref{circuit1})
represents the first two terms of the adiabatic expansion. In particular, it is easy 
to see that the second term is small as compared to the first one by the parameter
$\omega_0t_f\ll 1$. However, it accounts for the interaction effects, not contained 
in the first term.

\section{Boundary conditions and full counting statistics}
\label{boundary-c}

We have already mentioned, that it might be useful in a number of situations to 
express the bosonic fields $\phi_n$ in terms of their values away from the 
scattering region, where the statistics of their fluctuations is assumed to be 
known. Then, by solving equations of motion (\ref{eq-mot}) with corresponding boundary 
conditions, one can find correlation functions of the fields $\delta\phi_n$. This method
has been proposed in Ref.\ [\onlinecite{NB}] 
and successfully applied to a number of physical phenomena in quasi-1D systems far from 
equilibrium.\cite{experiments} Here we generalize this method to the case of arbitrary scattering. 

Let us assume that at the distance $W$ upstream the scattering region 
(see Fig.\ \ref{schematics1}) the fields $\delta\phi_n$ are known. According to 
the asymptotic form (\ref{boundary-cond}) of the scattering states, and 
taking into account the decomposition (\ref{fields2}), we can write
\begin{equation}
\delta\phi_n(-W,t)=
\int\limits_{0}^{\infty}\frac{d\omega}{\sqrt{\omega}}
\left[e^{-i\omega(W/v_n+t)}a_n(\omega)+\mbox{h.c.}\right].
\label{phi-W}
\end{equation}
On the other hand, according to Eq.\ (\ref{densities}),
the fluctuation of the charge injected into the system through the cross-section 
$x=-W$ is equal to 
\begin{equation}
\delta q_n(t)=\int\limits_{-\infty}^t dt'\delta j_n(-W, t')=-\frac{1}{2\pi}\delta \phi_n(-W,t).
\label{charges}
\end{equation}
Comparing these two equations, one finds that 
\begin{equation}
a_n(\omega)=-\sqrt{\omega}e^{i\omega W/v_n}\!\!\int\limits_{-\infty}^\infty
dt e^{i\omega t}\delta q_n(t),\quad \omega>0.
\label{an}
\end{equation}
By substituting this expression into Eq.\ (\ref{fields2}), we finally obtain
\begin{subequations}
\begin{eqnarray}
\delta \phi_n(x,t)=-\int\limits_{-\infty}^\infty dt'
\sum_m\Phi_{mn}(x, t-t')\delta q_m(t'),
\label{solution}\\
\Phi_{mn}(x,t-t')\equiv\int\limits_{-\infty}^\infty d\omega
\Phi_{mn\omega}(x)e^{-i\omega(t-t')}.
\label{set}
\end{eqnarray}
\label{results}
\end{subequations}
Here, we have used the property $[\Phi_{mn\omega}(x)]^*=\Phi_{mn-\omega}(x)$ and dropped the phase factor $e^{i\omega W/v_n}$, which merely shifts time (note, that the correlation functions for stationary processes do not depend on such time shifts). Omitting this phase factor simply amounts to redefining the scattering state. 

\begin{figure}[h]
\epsfxsize=7cm
\epsfbox{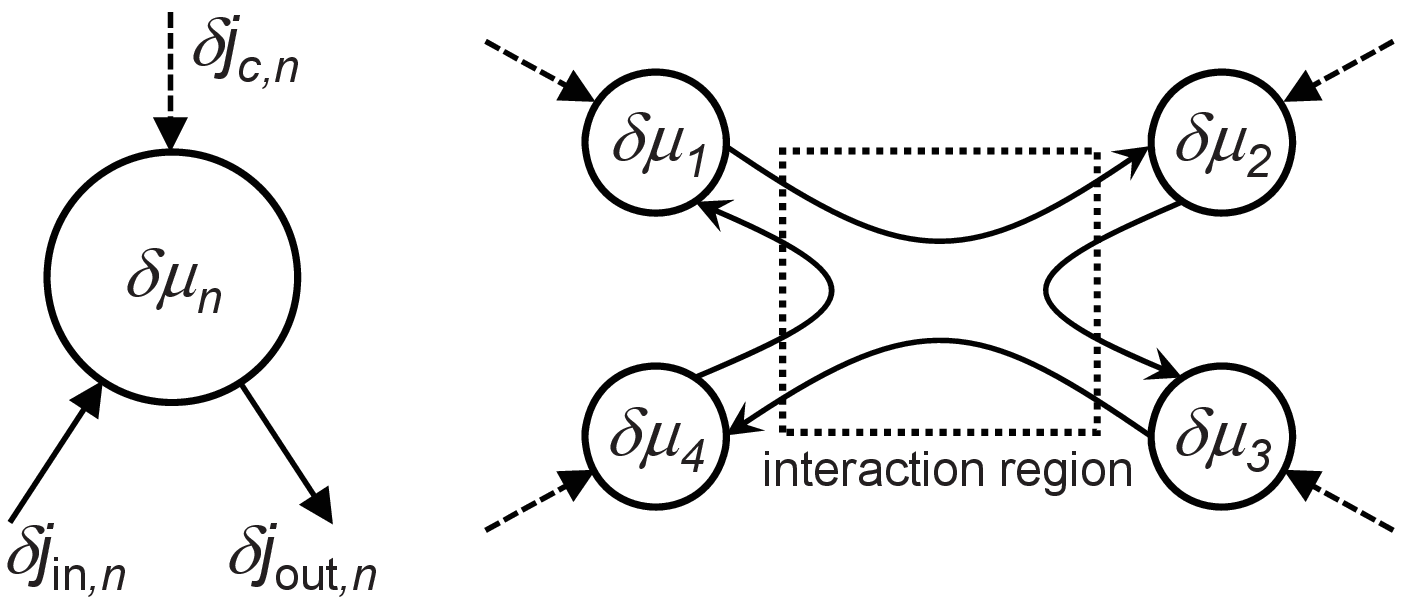}
\caption{
One of the reservoirs of electrons (an Ohmic contact) is schematically shown on the left. On one side, it is attached to an electrical circuit and receives the fluctuating current $\delta j_{c,n}$ from it, shown by the dashed line. This leads to fluctuations of the potential, $\delta\mu_n$. On the other side, it is connected to a quasi-1D electron system via two chiral channels carrying one incoming current $\delta j_{{\rm in},n}$ and one outgoing current $\delta j_{{\rm out},n}$. On the right, and example of a four-terminal systems is schematically shown.
} \vspace{-3mm}
\label{schematics3}
\end{figure}

We note that the statistics of the fluctuations of the fields $\phi_n(x,t)$, which are needed in order to calculate the fermion correlation functions, may be expressed, using Eqs.\ (\ref{results}), in terms of the statistics of fluctuations of the charges $q_n(t)$ transmitted through a given cross-section. Thus, the evaluation of fermionic correlation functions reduces to finding the full counting statistics of the transport of free fermions.\cite{FCS} Importantly, this method allows one to relax the assumption of the Gaussian character of fluctuations needed in order to arrive at the result (\ref{FR1}) and (\ref{FR2}).

\section{Langevin equations}
\label{LE}

In Secs.\ \ref{IB} and \ref{SS} we have considered conservative systems, while the results of Secs.\ \ref{ZM} and \ref{boundary-c} may also be used to account for the dissipation in electrical circuits, or, e.g.,
in tunnel junctions.\cite{NB} However, as far as circuit effects are concerned, it has been suggested earlier in Ref.\ [\onlinecite{us2}], that perhaps the most natural and efficient way to account for the dissipation is 
to apply the method of quantum Langevin equations. This method is based on the observation that electrical circuit elements typically create only Gaussian fluctuations, and the system in the bosonic sector remains Gaussian. Therefore, when accounting non-linear effects (such as weak tunneling or weak backscattering) perturbatively, one can describe the dynamics of the fields by using linear equations.

Let us consider $N$ electron reservoirs, in general at different temperatures $T_n$ and different potentials $\mu_n$. These reservoirs are connected, on one side, via an electrical circuit, characterized by the frequency-dependent conductance matrix $G_{nm}(\omega)$. On the other side, the reservoirs are attached to a quasi-1D electron system via chiral electron channels. This situation is illustrated in the Fig.\ \ref{schematics3}, where one of the reservoirs is schematically shown on the left. For simplicity only, and to illustrate our idea, let us assume, that each reservoir absorbs one incoming electron channel, and emits one outgoing channel (in the context of the QH effect this situation corresponds to the case of filling factor 1). 

We first concentrate on the fluctuation contribution to the fields (\ref{splitting}) and note that, according to the scattering theory [see Eq.\ (\ref{results})], it is determined by the currents  $\delta j_{{\rm out},n}$ outgoing from the reservoirs.
Thus, one can start with the equation for the charge conservation in the Fourier space,
\begin{equation}
-i\omega C_n\delta\mu_n=\delta j_{c,n}+\delta j_{{\rm in},n}-\delta j_{{\rm out},n},
\label{CC}
\end{equation}
where $C_n$ is the charge capacitance of the $n$th reservoir,\cite{footnote2} $\delta j_{c,n}$ is the fluctuation of the current incoming from the circuit, and the last two terms are the contributions of the electron channels.

The currents have contributions from the fluctuations of the collective modes, as well as from the Langevin sources. The current from the circuit acquires the following from:
\begin{equation}
\delta j_{c,n}=\sum_m G_{nm}\delta\mu_{m}+\delta j_{c,n}^s\,,
\label{circuit current}
\end{equation}
where $\delta j_{c,n}^s$ is the source.
Similarly, as it has been shown in Ref.\ [\onlinecite{us2}],
the outgoing current in the electron channel has two contributions,
 \begin{equation}
\delta j_{{\rm out},n}=G_q\delta\mu_{n}+\delta j_{{\rm out},n}^s\,,
\label{outgoing current}
\end{equation}
where $G_q=e^2/2\pi\hbar$ is the conductance quantum (restoring physical unites), and $\delta j_{{\rm out},n}^s$ is the equilibrium 1D current source originating from the reservoir. 

Finally, rewriting Eqs.\ (\ref{results}) in the frequency domain and in terms of currents injected to the system
\begin{equation}
\delta\phi_{n}(x,\omega)=-(2\pi i/\omega)\sum_m \Phi_{mn\omega}(x)\delta j_{{\rm out},m}(\omega)
\label{fluctuatinf-phi}
\end{equation}
and using Eq.\ (\ref{boundary-cond}), one connects incoming and outgoing 1D currents
\begin{equation}
\delta j_{{\rm in},n}(\omega)=\sum_mS_{nm}(\omega)\delta j_{{\rm out},m}(\omega),
\label{in-out}
\end{equation}
where $\delta j_{{\rm in},n}(-\omega)=\delta j_{{\rm in},n}^\dagger (\omega)$. These equations complete the set of the equations that should be solved for the currents $\delta j_{{\rm out},n}$ in terms of the sources. Knowing currents $\delta j_{{\rm out},n}$, one can find the fluctuating part of the fields $\delta\phi_n$ using Eq.\ (\ref{fluctuatinf-phi}). However, this has to be done with caution, because the matrix of conductances is degenerate. One can simply assume that one of the electron reservoirs is grounded, and corresponding potential does not fluctuate, or, alternatively, one may add an extra grounded electrode.

Since fluctuations are assumed Gaussian, it remains to accompany these results with the two-point correlation functions of the sources. It is natural to assume, that the sources are at local equilibrium, and correlation functions satisfy fluctuation dissipation relations:
\begin{equation}
\langle \delta j_{{\rm out},n}^s(\omega)\delta j_{{\rm out},m}^s(\omega')\rangle= \delta_{nm}\delta(\omega+\omega')\, 
\frac{2\pi\omega G_q}{1-e^{-\omega/T_n}}
\label{FDT1}
\end{equation}
and
\begin{equation}
\langle \delta j_{c,n}^s(\omega)\delta j_{c,m}^s(\omega')\rangle=
\delta(\omega+\omega')\, 
\frac{4\pi\omega G_{nm}}{1-e^{-\omega/T}},
\label{FDT2}
\end{equation}
where because of the chirality of 1D electrons
the factor of 2  in the first equation is missing as compared to the second one,
and in the second equation we assumed that the circuit is at equilibrium with the bath at the temperature $T$. After the fields are expressed in terms of the sources, one can use Eqs.\ (\ref{FDT1}) and (\ref{FDT2}) to find the correlators of the fields (\ref{FR2}), and eventially, the electron correlation functions (\ref{FR1}).

We conclude this section by addressing  the zero mode contributions $\varphi_n(x,t)$ to the fields $\phi_n(x,t)$. It has been emphasized in Sec.\ \ref{BofF} that the components $\varphi^{(0)}_n$ of zero modes guarantee the independence of fermions belonging to different 1D systems of finite size, and that in open quasi-1D systems this argument has to be reformulated. Here we propose to consider the whole system, i.e., the set of 1D fermion channels plus an electrical circuit, as an isolated ensemble of fermions, in which the number of particles does not change. This allows one to omit the phase factors $e^{i\varphi^{(0)}_n}$ in the bosonized representation of the fermionic operators, because fermions from different channels belong to the same system, are mixed (scattered) in the electrical circuit, and therefore are not strictly independent. Nevertheless, in practice their statistical independence  emerges from the fact, that different channels are connected to each other via reservoirs, where the phase fluctuations are strong. For example, if two fermionic channels are connected via one reservoir,  the corresponding correlator behaves as $\ln\langle \psi_n\psi_{n'}^\dagger\rangle\propto -T_nt_d$, where $T_n$ is the temperature, and $t_d$ is the dwell time of the fermion in the reservoir. Thus, this correlator vanishes in the thermodynamic limit.

The remaining components of the zero modes can be found by using the equations (\ref{0-modes2}) and (\ref{char}). According to the equation (\ref{0-modes2}), the zero modes are determined by the potentials of the reservoirs $\mu_n$, which in turn simply follow from the Kirchhoff's law, 
\begin{equation}
\langle j_{c,n}\rangle+\langle j_{{\rm in},n}\rangle - \langle j_{{\rm out},n}\rangle=0.
\end{equation} 
These equations have to be accompanied by the dc versions of the equations (\ref{circuit current}), (\ref{outgoing current}), and (\ref{in-out}): 
\begin{multline}
\langle j_{c,n}\rangle=\sum_m G_{nm}(0)\mu_{m},\quad \langle j_{{\rm out},n}\rangle=G_q\mu_{n},\\
\langle j_{{\rm in},n}\rangle=\sum_mS_{nm}(0)\langle j_{{\rm out},m}\rangle
\end{multline}
and solved for $\mu_n$. Here, in the notations of this section, $S_{nm}(0)=0,1$ is simply a connectivity matrix, because plasmons do not scatter at zero frequency.

\section{Discussion}

In the end, we would like to make important remarks and give practical recommendations concerning the application of the scattering theory. First of all, although we consider chiral systems throughout the paper, this is done for the convenience in order to simplify the derivations. With some limitations\cite{footnote0} on otherwise quite broad class of long-range interactions, our approach also applies to non-chiral systems, such as Luttinger liquids connected to free-fermionic reservoirs. In this case, the reservoirs are modelled by gradually switching off the interaction at the interface with the Luttinger liquid, which leads to the Andreev-type process, restoring the universality of the conductance of such systems.\cite{Ines1} Therefore, the interaction remains localized in a region of finite size, while incoming and outgoing states are still free. This is exactly the situation, where our method applies. Alternatively, one can simply change the basis of scattering states to the left and right movers and consider the interfaces with reservoirs as additional scattering centers. 

Second, in QH systems at integer filling factors larger than 1, co-propagating electron channels interact even in the asymptotically remote regions. However, this interaction is homogeneous and can be easily diagonalized by applying rotations of the basis in each sector of incoming states belonging to the same edge. Thus, the bosonic fields appearing in the vertex operators (\ref{def}) can still be expressed in terms of actual scattering states.

Third, additional electrostatic potentials, $V_n(x)$, e.g., induced locally by metallic gates, can easily be accounted by adding linear terms $(1/2\pi)\sum_n\int dx V_n\partial_x\phi_n$ to the Hamiltonian $H_0+H_1$. They slightly modify the equations of motion (\ref{eq-mot}) by adding the term $-V_n(x)$ to the right hand side. Consequently, this term arises in the right hand side of Eq.\ (\ref{0-eq-mot}) for zero modes, modifying the corresponding electrostatic problem, and shifting densities: $(1/2\pi)\partial_x\phi_n\to(1/2\pi)\partial_x\phi_n+\Delta \rho_n$. In turn, this introduces additional phase shift $2\pi\int dx\Delta\rho_n$ in fermionic operators (\ref{def}), in agreement with the Friedel sum rule.

Finally, we note that it would be interesting to extend our scattering theory in the analogy to the Floquet theory\cite{Floquet1,Floquet2} in order to investigate the photon-assisted electron transport in quasi-1D electron systems. This seems to be an obvious and straightforward next step, because even if a system is biased with time-dependent potentials, it remains Gaussian, and therefore the bosons are free.

\begin{acknowledgments}
We thank Ivan Levkivskyi and Artur Slobodeniuk for fruitful discussions.
This work has been supported by the Swiss National Science Foundation.
\end{acknowledgments}


\begin{thebibliography}{99}
\bibitem{footnote0}
This includes interactions leading to electron backscattering, such as Umklapp processes, 
which have to be considered perturbatively.
\bibitem{footnote1}
Strictly speaking, the prefactors in Eq.\ (\ref{def}) should also contain Klein factors,\cite{Thierry} which guarantee correct commutation relations for  fermions from different channels. This may become important if one needs to account high-order tunnelling processes.
\bibitem{footnote2}
In general, one may need to replace the left hand side of the equation (\ref{CC}) with $-i\omega\sum_m C_{nm}\delta\mu_m$, where $C_{nm}$ is the capacitance matrix that takes into account the Coulomb interaction between reservoirs.
\end{thebibliography}
\end{document}